\documentclass{acm_proc_article-sp}

\newcommand{\eat}[1]{}
\newcommand{\tg}{{\mbox{TGAP}}}
\newcommand{\Ot}{{\tilde{O}}}

\newcommand{\ghat}{{\hat{g}}}
\newcommand{\xhat}{{\hat{x}}}
\newcommand{\Lhat}{{\hat{L}}}
\newcommand{\what}{{\hat{w}}}
\newcommand{\xihat}{{\hat{\xi}}}
\newcommand{\gt}{{\tilde{g}}}
\newcommand{\gb}{{\overline{g}}}

\newtheorem{theorem}{Theorem}
\newtheorem{lemma}{Lemma}

\begin{document}

\title{Duality and Nonlinear Graph Laplacians}
%
%
%
%
%

\numberofauthors{2} 
%
\author{
%
%
\alignauthor
Eric J. Friedman\\
       \affaddr{International Computer Science Institute}\\
       \email{ejf@icsi.berkeley.edu}
\alignauthor
Adam S. Landsberg\\
       \affaddr{W. M. Keck Science Department of Claremont McKenna, Pitzer, and Scripps Colleges}\\
       \email{ALandsberg@kecksci.claremont.edu }
}

\maketitle

\begin{abstract}
We present an iterative algorithm for solving a class of \\nonlinear Laplacian system of equations in  $\tilde{O}(k^2m \log(kn/\epsilon))$ iterations, where $k$ is a measure of nonlinearity, $n$ is the number of variables, $m$ is the number of nonzero entries in the graph Laplacian $L$, $\epsilon$ is the solution accuracy and $\tilde{O}()$ neglects (non-leading) logarithmic terms. This algorithm is a natural nonlinear extension of the one by of Kelner et. al., which solves a linear Laplacian system of equations in nearly linear time. Unlike the linear case, in the nonlinear case each iteration takes $\tilde{O}(n)$ time so the total running time is $\tilde{O}(k^2mn \log(kn/\epsilon))$. For sparse graphs where $m = O(n)$ and fixed $k$ this nonlinear algorithm is $\tilde{O}(n^2 \log(n/\epsilon))$ which is slightly faster than standard methods for solving linear equations, which require approximately $O(n^{2.38})$ time. Our analysis relies on the construction of a nonlinear ``energy function'' and a nonlinear extension of the duality analysis of Kelner et. al  to the nonlinear case without any  explicit references to spectral analysis or electrical flows.   These new insights and results provide tools for more general extensions to spectral theory and nonlinear applications.
\end{abstract}


\terms{Graph Laplacian, Spectral analysis, Nonlinear equations}


\section{Introduction}

The goal of solving linear systems of equations in nearly linear time, spearheaded by Vaidya \cite{vaidya1991solving}, has propelled a revolution in spectral graph theory and several related fields.  In addition to many important results about graph sparsification (see \cite{spielman2010algorithms,vishnoi2012laplacian} for overviews and references) this theory has led to two main types of methods of solving sets of symmetrical diagonally dominant linear equations in nearly linear time -- the first using graph sparsification (and ultra sparsifiers) (e.g., \cite{spielman2004nearly}) combined with classical iterative methods, and the second which relies on a underlying low stretch spanning tree to provide an algorithm that relies on cycle updates for ``electrical flows'' in the underlying graph \cite{kelner2013simple}.  In addition to having  important applications in matrix computations, these results have led to a large number theoretical results, reducing the complexity of several important graphical algorithms.

The key insights underlying these approaches are the reduction of a set of symmetrical diagonally dominant linear equations to a system of equations $Lx=b$ arising from a graph Laplacian, $L$.
Previous analyses (e.g., \cite{spielman2004nearly}) rely on the linearity of the underlying system of equations and the use of a carefully chosen spanning tree on the underlying graph.  Thus, one might expect that linearity is required for the analyses; however, as we will show, linearity is not necessary and there appears to be a more general structure underlying these approaches which allows a certain degree of nonlinearity.  More precisely, we show that one can generalize the graph Laplacian in a nonlinear manner and  apply a nonlinear extension of the Algorithm in Kelner et al. \cite{kelner2013simple} to solve them in  $\Ot(k^2m \log(k n/\epsilon))$ iterations and $\Ot(k^2mn \log(k n/\epsilon))$ time.  Our analysis relies on the construction of an ``nonlinear energy function'' and extension of  their duality analysis to  nonlinear Laplacian systems of equations. While providing new insights into the linear problem, this analysis shows that one can generalize graph Laplacians without losing much of the underlying structure and suggests further nonlinear generalizations.

\section{Related Work}

The idea of using a spanning tree to precondition a set of linear equations comes from an unpublished presentation by Vaidya \cite{vaidya1991solving}.  That paper led to a large body of work over many years that led to Spielman and Teng's seminal  paper \cite{spielman2004nearly} which showed the theoretical existence of nearly linear time algorithms for systems of equations, using very sophisticated machinery \cite{koutis2012fast,reif1998efficient,spielman2011graph,spielman2009note}.  

Our algorithm is based on Kelner et. al's algorithm \cite{kelner2013simple}, which dramatically reduces the amount of machinery required. Its origins are clearly implicitly based on the more sophisticated machinery in the earlier papers, such as low-stretch spanning trees (e.g., \cite{abraham2012using} ) and ultra-sparsifiers  (e.g., \cite{spielman2004nearly}), but does not require them directly.

\section{Generalized Model and Relaxation }

We begin with an undirected weighted graph $G= <V,E,w>$ with node set $V$, edges $E$ and weights $w_{ij}>0$ where  $w_{ij}=w_{ji}$.   The graph Laplacian of $G$ is given by $L$ where $L_{ij}= -w_{ij}$ for $i\not = j$ and $L_{ii}=-\sum_{j\not=i} w_{ij}$.  Given a vector $b\in \Re^n$ a Laplacian system of equations is simply $Lx =b$. Such a system always has a solution if $\sum_i b_i = 0$, which we denote $x^*$ and is unique up to an additive constant.  Throughout this paper whenever we talk about solutions, we will ignore the multiplicity of solutions due to the nullspace of $L$. Note that for notational clarity we will always assume that $G$ is connected. In the case that $G$ is not connected one can simply solve  on each connected component separately.

Let $N(i)$ be the neighbors of node $i$ in the graph $G$.  It is well known that one can express the action of the Laplacian as 
$$(Lx )_i = \sum_{j\in N(i)} w_{ij} (x_i-x_j),$$ which only depends on $x_i-x_j$ for $(i,j)\in E$.
In the following we will consider a generalization of this  to non-linear graph Laplacians where 
$$(Lx )_i = \sum_{j\in N(i)} w_{ij} h_{ij}(x_i-x_j)~~~~~~~(*)$$ 
where the nonlinear function $h_{ij}(\cdot)$ is anti-symmetric, satisfies $h_{ij}(-v) = -h_{ij}(v)$,  is continuous, the derivative $h'_{ij}(v)$ is left-continuous, $v h_{ij}'(v)$ is strictly increasing and $1/k \leq h_{ij}'(v)<k$ for a chosen $k\geq 1$. 

Note that the running time of our algorithm will depend on $k$ and the computational complexity of $h_{ij}(\cdot)$. In order to simplify the analysis with respect to $h_{ij}(\cdot)$, we assume that we have constant time oracles for computing $h_{ij}(\cdot)$, $h_{ij}^{-1}(\cdot)$, and $h_{ij}'(\cdot)$.  The more natural version where these oracles require $O(\log(1/\epsilon))$ computations, would only increase our running times by $O(\log(1/\epsilon))$ factors, where $\epsilon$ is the accuracy of the solution.

A simple example of this model arises when $h_{ij}'(v) = 1/k$ for $|v|\leq 1$ and $h_{ij}'(v) = 1$ otherwise.  This example arises when the cost is small for small flows, in contrast to the linear case where the ``cost'' of a flow between edges is linear in the flow. More generally, we can choose  $h_{ij}'(v)$ to be constant except at a discrete set of points where it increases, corresponding to a piecewise linear $h_{ij}(v)$ with discrete increases in slope, bounded by slopes $1/k$ and $k$.  Surprisingly, one can also use an $h_{ij}(v)$ that is not convex in the positive quadrant. For example the function given by $h_{ij}(v) = v+arctan(v)$  satisfies the requirements and is not convex in the positive quadrant, leading to``economies of scale'' where the incremental cost can decrease as a flow increases.

Following Kelner et. al \cite{kelner2013simple} we can introduce new variables,  $g_{ij}$, which  for generalized Laplacians allow us to rewrite $Lx=b$ as:
$$ \forall i\in V:~~\sum_{j\in N(i)} w_{ij} h_{ij}(g_{ij}) = b_i ~~~~(BC_i)$$
$$\forall (i,j)\in E:~~~g_{ij} = x_i-x_j  ~~~~(PC_i)$$
which decomposes the problem in a natural way into b-constraints ($BC_i$) and p-constraints ($PC_i$), where  $x$ is a `potential function' for  $g$. We say that $g$ is b-feasible if $g$ satisfies the b-constraints and p-feasible for p-constraints.  

Formally, we will only have a single variable for each edge, written $g_e$ for $e\in E$ but for notational convenience we will write  $g_{ij}=-g_{ji}$ where one of these two variables as canonical and the other is simply a notational convenience.  For example  if $\sum_{(i,j)\in E} g_{ij}$ corresponds to a sum where each edge appears once, while in $\sum_{i\in V} \sum_{j\in N(i)}  g_{ij}$ each $g_{e}$ appears twice. 

Note that our variables differ from those in Kelner et. al \cite{kelner2013simple} and most of the literature  in the linear case where in the electrical model our $g_{ij}$ variables correspond to the voltage drop between nodes $i$ and $j$ while their $h_{ij}$ corresponds to the current flow between those nodes. Algebraically, $g_{ij} = h_{ij}/w_{ij}$ .

\section{Cycle Updates}

A key idea in the algorithm by Kelner et al \cite{kelner2013simple} is to update the $g_{ij}$'s using cycles to maintain b-feasibility.  Let $C$ be an oriented set of edges in $G$ that forms a cycle.  Let $\alpha(C,t)$ be defined such that for all $(i,j)\in C$,  
$$w_{ij} h_{ij}(g_{ij}+\alpha_{ij}(C,t)) = w_{ij} h_{ij}(g_{ij}) +t,$$
and $\alpha_{ij}(C,t)=0$ for $(i,j)\not\in C$ and note that $\alpha(C,0)=0$ by construction.
It can be easily checked  that this procedure maintains b-feasibility for any value of $t$, since the increase from an incoming edge of the cycle through any node is canceled by the decrease in the sum from the outgoing edge that follows it. When $h_{ij}(\cdot)$ is the identity,   $\alpha_{ij}(C,t) = t/w_{ij}$ as in the linear analysis.

The key idea in the linear analysis is to iteratively update cycles by updating cycles using $\alpha(C,t)$ in order to approach p-feasibility while always maintaining b-feasibility. The analysis  \cite{kelner2013simple} relies on duality related to ``electrical systems'' with an  energy function. In the nonlinear case  we can also define an analogous energy function 
$$\Phi(g)= \sum_{ij}\phi_{ij}(g_{ij})$$  where $g$ is the vector of $g_{ij}$'s and 
$$\phi_{ij}(g_{ij}) = w_{ij} \int_0^{g_{ij}}s h_{ij}'(s)~ ds.$$
In the linear case  where $h_{ij}(v)$ is the identity, this reduces to the standard energy function of \cite{kelner2013simple}, after a change of variables. 

The key insight in our analysis is that for any cycle $C$ the ``cycle adjustment'', $\alpha(C,t)$ that minimizes $\Phi(g)$ satisfies p-feasibility around the cycle.

\begin{lemma}\label{lemma:cycle}
If for some cycle $C$ 
$$t^* = \mbox{argmin}_t \Phi(g+\alpha(C,t))$$
then
 $$\sum_{(i,j)\in C } (g_{ij}+\alpha_{ij}(C,t^*)) =0.$$
 \end{lemma}
 \proof
To see this note that $\Phi(g)$ is convex by the definition of the $h$'s and that the derivative is
$$ \frac{d\Phi(g+\alpha(C,t))}{dt}=\sum_{(i,j)\in C} \frac{d\phi_{ij}(g_{ij}+\alpha_{ij}(C,t))}{dt} $$  
$$~~~~~~~~~~~~~~~~~~~= \sum_{(i,j)\in C} \phi_{ij}'(g_{ij}+\alpha_{ij}(C,t))\frac{d\alpha_{ij}(C,t)}{dt}.$$
Inserting the definition of $\phi_{ij}$ into the right hand side of the equation yields 
$$~~~~~~~~~~~ = \sum_{(i,j)\in C} w_{ij} (g_{ij}+\alpha_{ij}(C,t))h_{ij}'(g_{ij}+\alpha_{ij}(C,t))\frac{d\alpha_{ij}(C,t)}{dt}.$$
Implicitly differentiating the defining equation for $\alpha(C,t)$ with respect to $t$ gives
$$w_{ij} h_{ij}'(g_{ij}+\alpha_{ij}(C,t)) \frac{d\alpha_{ij}(C,t)}{dt}= 1 , ~~(**)$$
which combined with the previous equation and the definition of $t^*$ gives the desired result after  setting the derivative to 0.~$\Box$

This lemma implies our first theorem:
 
\begin{theorem}
If $g$ is a b-feasible flow that minimizes $\Phi(g)$ over all ``cycle updates'' in $G$, then $g$ is also p-feasible.
\end{theorem}
\proof
Clearly if there are no cycle updates that decrease the energy then by the preceding Lemma the flows around those cycles must sum to $0$.~$\Box$

We next derive a quantitative measure of the gain in energy from a cycle update. This will depend on both the initial flow around the cycle $$G_C= \sum_{(i,j)\in C} g_{ij}$$ and the harmonic sum of the $w$'s around the cycle $$W_C= (\sum_{(i,j)\in C} 1/w_{ij})^{-1}.$$
\begin{lemma}
For a cycle $C$ the decrease in energy from a cycle update satisfies:
$$\Phi(g)-\Phi(g+\alpha(C,t^*)) \geq  W_C G_{C}^2/(2k).$$
\end{lemma}
\proof
We assume that $G_C \leq 0$. (If not then simply traverse the cycle in the opposite direction.)
As seen in the proof in Lemma~\ref{lemma:cycle}
$$ \frac{d\Phi(g+\alpha(C,t))}{dt}= \sum_{(i,j)\in C} (g_{ij}+\alpha_{ij}(C,t)).$$
Integrating from $t=t^*$ to $t=0$ yields
$$ \Phi(g)-\Phi(g+\alpha(C,t^*))= \int^0_{t^*} (G_C +\sum_{(i,j)\in C} \alpha_{ij}(C,t))dt~~~~(***).$$
Note that the integrand  is $0$ at $t=t^*$, $G_C$ at $t=0$ and is strictly decreasing over the range of integration.  
Using (**)  we see that
$$ \frac{d\alpha_{ij}(C,t)}{dt}= (w_{ij} h_{ij}'(g_{ij}+\alpha_{ij}(C,t)))^{-1}\leq k/w_{ij}$$
Summing over $C$ and integrating over $t$ yields
$$ \sum_{(i,j)\in C} \alpha_{ij}(C,t) \leq kt/W_C$$
Since $G_C+ \sum_{(i,j)\in C} \alpha_{ij}(C,t^*) =0$ this
implies that $t^* \geq -W_C G_C/k$. Combining this with (***) yields
$$ \Phi(g)-\Phi(g+\alpha(C,t^*)) \geq \int^0_{ -W_C G_C/k} (G_C +kt/W_C)dt$$
and integrating completes the proof.~$\Box$

To compute  the cycle update we can use Newton's method on the derivative $\Phi( g+\alpha(C,t))$, written in the form. 
$$ G_C+\sum_{(i,j)\in C} \alpha_{ij}(C,t) .$$
Note that the sum is bounded by $(t/W_C)/k$ and $(t/W_C)k$  so we can use a binary search on the interval $[0,(t/W_C)k]$.  Since the function is convex in $t$, for any $t\in[t^*/4,3t^*/4]$ the value of the function will be reduced by at least a factor of 2. This will require $O(\log(k^2 ))$ iterations so the total time required is  $O(n \log(k))$ since each evaluation of the  sum requires $O(n)$ time.  

Thus we have proven an approximate version of the previous lemma.

\begin{lemma}\label{lemma:update}
One can compute an approximate cycle update in $O(n \log(k))$  such that for any cycle $C$ the decrease in  energy satisfies:
$$\Phi(g)-\Phi(g+\alpha(C,t^*)) \geq  W_C G_{C}^2/(4k).$$
\end{lemma}

\section{Spanning Trees}

Given a spanning tree, $T\subseteq E$ one can analyze a simple set of cycles that span the cycle space of all cycles \cite{bollobas1998modern}. Given an edge $(i,j)\in E\setminus T$ define the tree path $P_{(i,j)}$ to be the minimal  oriented set of edges from node $j$ to node $i$ and let the tree cycle to be  $C_{(i,j)}=P_{(i,j)}\bigcup (i,j)$.  Thus, instead of minimizing $\Phi(g)$ over all cycles one only needs to minimize it over each of the tree cycles.

As in \cite{kelner2013simple}, we will  choose edges $e\in E\setminus T$ at random with probability $p_e$ and then minimize $\Phi(g)$ over the tree cycle $C_e$.  The main difference here is that in the linear case one can minimize the energy over a cycle analytically while in the nonlinear case one needs to rely on an iterative method, which we described earlier.  

For later use, we define the tree-values of $x$, denoted $\xhat(T,g)$ to be those which satisfy $g_{ij}=x_i-x_j$ on the tree. These are easily computed, up to an additive constant, using using a breadth first search  from any leaf of the spanning tree of $G$.

\section{Lagrangian Duality}

In order to analyze the progress made by each iteration of our algorithm we need an estimate of the distance to optimality. To do this we compute the Lagrangian dual for our minimization problem.

First recall that our primal problem is to compute 
$$\min_g \Phi(g)~~~s.t.~~~~\sum_{j\in N(i)} w_{ij}h_{ij}(g_{ij}) = b_i.~~~\forall i\in V $$
The dual function is given by 
$$\Theta(x) = \min_g \{\Phi(g) - \sum_{ i\in V} x_i [( \sum_{j\in N(i)} w_{ij}h_{ij}(g_{ij})) - b_i]\}$$
where $x$ are the dual variables, and as we will see correspond to the $x$ in our original problem.
(As a reminder, we only consider one of $g_{ij},g_{ji}$ as a variable, say $g_{ij}$ and implicitly replace all occurrences of $g_{ji}$ with $-g_{ij}$.)
The first order condition for this minimization when differentiating by $g_{ij}$ is: 
$$0  = \phi_{ij}'(g_{ij})-(x_i -x_j) w_{ij}h_{ij}'(g_{ij})$$
Using the definition of $\phi_{ij}$ yields:
$$0  = w_{ij}h_{ij}'(g_{ij})g_{ij}-(x_i -x_j) w_{ij}h_{ij}'(g_{ij})$$
which reduces to $g_{ij}=x_i-x_j$.
Plugging this into $\Theta(x)$ yields
$$\Theta(x) = \sum_{(i,j)\in E} \phi_{ij}(x_i-x_j)-\sum_{ i\in V} x_i [(\sum_{j\in N(i)} w_{ij}h_{ij}(x_i-x_j) )- b_i]$$
To simplify this we define $\gt_{ij} = x_i-x_j$ and use the b-feasibility of $g_{ij}$ to write this as
$$\Theta(x) = \sum_{(i,j)\in E} \phi_{ij}(\gt_{ij})-\sum_{(i,j)\in E} w_{ij}\gt_{ij}(h_{ij}(\gt_{ij})- h_{ij}(g_{ij})).$$

Using this we can approximate the distance to optimality since $\Phi(g)\geq\Theta(x)$ for any b-feasible $g$ and $\Phi(g^*)=\Theta(x^*)$ where $g^*,x^*$ are the optimizers of their respective problems.  Thus, following Kelner et. al if we define $$GAP(g) = \Phi(g)-\Phi(g^*)$$ then for any b-feasible $g$ we have
$$GAP(g) \leq \Phi(g)-\Theta(x)$$
We can use the tree values for $x$ to provide this estimate yielding $$GAP(g) \leq \tg = \sum_{(i,j)\in E\setminus T}\tg_{ij}(g)$$ where we define $\tg$ by
$$\tg_{ij}(g)= \phi_{ij}(g_{ij})-\phi_{ij}(\xhat_i-\xhat_j)~~~~~~~~~~~~~~~~~~~~\mbox{}$$
$$\mbox{}~~~~~~~+w_{ij} (\xhat_i-\xhat_j)(h_{ij}(\xhat_i-\xhat_j)-h_{ij}(g_{ij}))$$
where  the sums in the gap formula need only be over non-tree edges since $x_i-x_j=g_{ij}$ for all $(i,j)\in T$.

\begin{lemma}\label{lemma:gap}
Given a b-feasible $g$, $$GAP(g) \leq \sum_{(i,j)\in E\setminus T}  w_{ij} G_{C_{ij}}^2 k /2$$
where $C_{ij}$ is the tree cycle passing through edge $(i,j)$.
\end{lemma}
\proof
Since $GAP\leq TGAP$ we can focus on TGAP,
$$\frac{\partial \tg_{ij}(g)}{\partial g_{ij}} = \phi_{ij}'(g_{ij}) - w_{ij}(\xhat_i-\xhat_j)h_{ij}'(g_{ij})$$
substituting for $\phi_{ij}'$ gives
$$\frac{\partial \tg_{ij}(g)}{\partial g_{ij}} = w_{ij}(g_{ij} -  (\xhat_i-\xhat_j)) h_{ij}'(g_{ij}).$$
Integrating  yields
$$ \tg_{ij}(g)\leq   \int_{(\xhat_i-\xhat_j)}^{g_{ij}} w_{ij} (s_{ij} -  (\xhat_i-\xhat_j)) k ~ds_{ij} $$
which after noting that $G_{C_{ij}}= (g_{ij} -  (\xhat_i-\xhat_j))$ yields
$$ \tg_{ij}(g)\leq   w_{ij} G_{C_{ij}}^2 k $$
which implies the result after summing.
~$\Box$

\section{Tree Condition Numbers}

In order to guarantee convergence in the linear case, Kelner et. al. \cite{kelner2013simple} require that the underlying spanning tree have nice properties. As we will show the same condition suffices for the nonlinear case as well.  

Define  the tree condition number of spanning tree T to be 
$$ \tau(T) =  \sum_{(i,j)\in E\setminus T}  w_{ij}/W_{C_{ij}} $$
where $C_{ij}$ is the tree cycle over edge $(i,j)$.
Now, as discussed in Kelner et. al, the tree condition number is closely related to the stretch of a tree, a well studied topic.  
The stretch of a tree $T$ is defined to be
$st(T) = \sum_{(i,j)\in E} st(i,j)$ where 
$$st(i,j) = w_{ij} \sum_{(r,s)\in P_{(i,j)}} 1/w_{rs}$$
and it is easy to show that $\tau(T)=st(T)+m-2n+2.$

Thus, we can use the following result due to Abraham et. al \cite{abraham2012using} to construct a tree with a good condition number

\begin{theorem}[Abraham et. al \cite{abraham2012using}] \label{thm:stretch}
In $O(m \log n \log \log n)$ time we can compute a spanning tree $T$ with total stretch \\
$st(T) = O(m \log n\log\log n).$
\end{theorem}

\section{ Linearization}

For static analyses we can linearize the Laplacian at some $\ghat$ by considering the adjusted weights, $\what^\ghat_{ij} = w_{ij} h_{ij}(\ghat_{ij})/\ghat_{ij}$, creating a new Laplacian $\Lhat^\ghat$. Then we can write
$$(\Lhat^\ghat x)_i = \sum_{j\in N(i)} \what^\ghat_{ij} (x_i-x_j).$$
Next  we define  the linearized energy to be  the energy of the linearized Laplacian:
$$\xihat^\ghat(g) = \sum_{(i,j)\in E} \what_{ij}^\ghat g_{ij}^2/2$$
and it is easy to see that $\xihat^\ghat(\ghat) = \Phi(\ghat)$.
Then, using the bounds on $h'_{ij}(g_{ij})$ we see that
$$\what^\ghat_{ij}  \in [(1/k)w_{ij},kw_{ij}],$$
$$\Phi(g) \in [(1/k)\xihat^\ghat(g),k\xihat^\ghat(g)]$$
and 
$$\Phi(g)/\Phi(g^*) \geq (1/k^2) \xihat^\ghat(g)/\xihat^\ghat(g^*).$$

This will allow us to use results from the linear case to analyze some static properties of our algorithm.

\section{Accuracy}

In the linear case, most spectral methods  consider the accuracy in solution  \cite{kelner2013simple}. Thus we will say that a solution is $\epsilon$-accurate if   
$$\frac{||x-x^*||_{L^{q^*}}}{||x^*||_L^{q^*}}\leq \epsilon,$$
 where $L^{q^*}$ is the linearization of $L$ at $q^*$, as used in the previous section.  
By  combining our bounds on the linearization of the energy function (***) with Lemma~6.2 in \cite{kelner2013simple} for the linearization we can easily prove the following:

\begin{lemma}\label{lemma:final}
If $$\Phi(g)/\Phi(g^*)\leq 1+\epsilon^2/(\tau(T) k^4)$$ then  $$\frac{||x-x^*||_L^{q^*}}{||x^*||_L^{q^*}} \leq \epsilon $$
where $x$ is defined by its  tree values on $T$.
\end{lemma}
\proof
It is easy to see from Lemma~6.2 in \cite{kelner2013simple} that 
if $\Phi(g)/\Phi(g^*)\leq (1+\epsilon^2/k^4)$ then  $||x-x^*||_{\Lhat^g} \leq \epsilon/(k^2\tau(T))$. To complete the proof we need the following lemma relating 
$||v||_{\Lhat^g}$ and 
$||v||_{\Lhat^{g^*}}$.~$\Box$

\begin{lemma}
$||v||_{\Lhat^g} \leq k^2 ||v||_{\Lhat^{g^*}}$.
\end{lemma}
\proof
This follows from the definition:
$$ ||v||_{\Lhat(g)} = \sum_{(i,j)\in E} \what^g (v_i-v_j)^2$$
and noting that $\what_{ij}^g \in [(1/k)w_{ij},kw_{ij}]$.~$\Box$

\section{Initialization}

To initiate our algorithm we start with a b-feasible flow $g^0$ with support only on the spanning tree.  This can be computed recursively with $O(n)$ operations by noting that for any leaf node $i$ of the tree the value of $g_{ij}$ for $j=N(i)$ is given by $w_{ij} h_{ij}(g_{ij}) = b_i$.

However, it will be useful to note that this can be computed by first computing the solution for the linearization of $L$ denoted  $\gb^0$, where we set  all $h_{ij}(\cdot)$ to be identity functions, using the above recursive algorithm.  Then  convert $\gb^0$ to $g^0$ using the relationship $g^0_{ij} = h_{ij}^{-1}(\gb^0_{ij})$. 

The following bound on the initial energy follows directly from Lemma~6.1 in \cite{kelner2013simple}.
\begin{lemma}\label{lemma:initial}
$\Phi(g^0) \leq k^2 \Phi(g^*) st(T)$
\end{lemma}
\proof
Lemma~6.1 in \cite{kelner2013simple} shows that $\xi(\gb^0) \leq k^2 \xi(\gb^*) st(T)$. The result follows from the fact that 
$$\Phi(g) \in [(1/k)\xihat^\ghat(g),k\xihat^\ghat(g)].$$~$\Box$

\section{Algorithm and Main Result}

We now present the algorithm which is a slight modification of that in Kelner et. al \cite{kelner2013simple}.

{\noindent \bf Algorithm: Simple Nonlinear Solver}
\begin{enumerate}
\item INPUT: Input $G=(V,E,w)$, $\epsilon\in \Re^+$, $k\geq 1$, $h_{ij}(\cdot)~~\forall (i,j)\in E$.
\item OUTPUT:  $g\in \Re^E$, $x\in \Re^V$.
\item Construct $T$: a low stretch spanning tree of $G$.
\item Construct $g^0$: a b-feasible solution with support only on the spanning tree.
\item Set probabilities $p_{ij}= w_{ij}/(W_{C_{ij}} \tau(T)) $ for all $(i,j)\in E\setminus T$.  
\item Set $S = \lceil  2 k^2 \tau (T) \log(k^6 st(T)\tau(T)/\epsilon^2) \rceil $.
\item For $\ell$ = 1 to S:
	\begin{enumerate}
	\item Pick a random $(i,j)\in E\setminus T$ with probability $p_{ij}$.
	\item Apply a cycle update to $g^{\ell-1}$ using tree cycle $C_{ij}$ to get $g^{\ell}$. Compute this to relative accuracy $1/(2k^2)$ using binary search.
	\end{enumerate}
\item RETURN $g$ and its tree induced $x$.
\end{enumerate}

In order to show the correctness of this algorithm, we extend the analysis in \cite{kelner2013simple}. Since the $GAP(g^\ell)$ conditional on $GAP(g^{\ell-1})$ is a random variable we focus on expected values.

\begin{lemma}\label{lemma:reduction}
For each iteration in the loop of ``Algorithm: Simple Nonlinear Solver'' 
$$\frac{E[GAP(g^\ell)] -E[GAP(g^{\ell+1})|GAP(g^{\ell})]}{E[GAP(g^{\ell})] } \geq \frac{1}{2 k^2 \tau(T)}.$$
\end{lemma}
\proof
First note that the change in GAP is the same as the change in  $\Phi(g)$ which is greater than $W_{C_{ij}} G_{C_{ij}}^2/(2k)$ for an update of $C_{ij}$, thus the expected decrease is 
$$E[GAP(g^\ell)]-E[GAP(g^{\ell+1})|GAP(g^{\ell})] \geq \sum_{(i,j)\in E\setminus T} p_{ij} W_{C_{ij}} G_{C_{ij}}^2/(2k).$$
Substituting in $p_{ij} = w_{ij}/(\tau(T) W_{C_{ij}})$ yields
$$E[GAP(g^\ell)]-E[GAP(g^{\ell+1})|GAP(g^{\ell})] \geq \sum_{(i,j)\in E\setminus T} (w_{ij}/\tau(T))G_{C_{ij}}^2/(2k)$$
using Lemma~\ref{lemma:gap} for the RHS completes the proof.~$\Box$

Using this we can prove our main theorem.
\begin{theorem}
Given a nonlinear graph Laplacian $L$, ``Algorithm: Simple Nonlinear Solver'', computes an expected $\epsilon$-accurate  solution of $Lx=b$ in
$\Ot(mnk^2 \log(k n/\epsilon))$  time.
\end{theorem}
\proof
Rewriting the expression in Lemma~\ref{lemma:final}  shows that in order to attain $\epsilon$-accuracy we need 
$$\Phi(g^S)-\Phi(g^*)\leq \epsilon^2/(\tau(T)k^4)\Phi(g^*).$$  
Iterating Lemma~\ref{lemma:reduction}, implies that 
$$\Phi(g^S)- \Phi(g^*) \leq (1-\frac{1}{2 k^2 \tau(T)})^S (\Phi(g^0)-\Phi(g^*)).$$
Combining this with Lemma~\ref{lemma:initial} on the initial energy yields
$$\Phi(g^S)- \Phi(g^*)\leq (1-\frac{1}{2 k^2 \tau(T)})^S (k^2  st(T)-1) \Phi(g^*).$$
Plugging in $S = \lceil  2 k^2 \tau (T) \log(k^6 st(T)\tau(T)/\epsilon^2) \rceil $ satisfies the requirement, proving the accuracy of the Algorithm.
By Theorem~\ref{thm:stretch} 
$$st(T) = O(m \log n\log\log n) = \Ot(m)$$ the required number of iterations is given by 
$$r = \Ot(k^2 m \log(k  n/\epsilon)).$$
Since each iteration takes $\Ot(n)$ time by Lemma~\ref{lemma:update} we get the stated result.~$\Box$

Note that instead of an expected $\epsilon$-accurate solution, one can  compute an actual $\epsilon$-accurate solution with high probability with the same time bound  using standard probabilistic techniques.

\section{Concluding Remarks}

This paper has shown that one can extend some of the techniques from the linear theory of systems of equations arising from linear graph Laplacians to ones with nonlinear graph Laplacians.  We view our analysis as suggestive that even more general extensions, perhaps even asymmetric ones related to directed graphs, may be possible, extending the range of  spectral analysis. Perhaps as in Kelner et. al \cite{kelner2013simple} our algorithm has a representation as an alternating projections algorithm, showing that nonlinear extensions are also possible in that setting.

There are many open directions for improving our algorithm. First, note that we have not optimized the $k$ dependence. Next, note that one could use the ideas of tree scaling and a randomized stopping time  as in \cite{kelner2013simple} to reduce the running time. In addition, one might be able to use  algorithms for the linear problem to speed up the nonlinear algorithm, either by providing a warm start or by speeding up convergence near optimality when the linear problem might provide a good approximation for the nonlinear one. 

While this paper raises many open problems relating to nonlinear spectral theory, the main problem with the current algorithm is the dependence on $n$ due to the complexity of computing a nonlinear cycle update which unlike the linear version does not have a simple analytic solution which can be computed in logarithmic time using a special data structure for keeping track of the flows on the tree. Nonetheless, it might be possible to reduce this complexity either through a specially designed data structure for the spanning tree, or perhaps through the use of a spanning tree with small diameter.  Also, the direct application of our algorithm to nonlinear graph problems might be fruitful.


\section{Acknowledgments}
This material is based upon work supported by the National Science Foundation under Grants No. 1216073 and 1161813. 

%
\bibliographystyle{abbrv}
\bibliography{../bib} 
%
%
\end{document}